\documentclass{article}
\usepackage{amsmath}
\usepackage{amsfonts}
\usepackage{epsfig}
\usepackage{rotating}
\usepackage{graphicx}
\usepackage{amssymb}
\usepackage{eucal}
\usepackage{lscape}
\usepackage{latexsym,enumerate}

\usepackage{fancyhdr}
\usepackage{natbib}
\usepackage{pdfpages}

\newtheorem{defn}{Definition}[section]

\newtheorem{theorem}[defn]{Theorem}
\newtheorem{cor}[defn]{Corollary}
\newtheorem{assumption}[defn]{Assumption}

\begin{document}

\title{Sex at birth could well be a biological coin toss....\\ Beware of conditioning on post-baseline information}

\author{Judith J.~Lok$^1$ and Mireille Schnitzer$^2$\\
$\;^1$Department of Mathematics and Statistics, Boston University\\
$\;^2$Faculty of Pharmacy, Universit\'{e} de Montr\'{e}al}

\maketitle

{\bf Abstract.} \cite{wang2025sex} use statistics to argue that sex at birth is not a biological coin toss, by noticing that repeated patterns such as Male Male Male and Female Female Female occur in the Nurses Health Study more often than patterns like Male Female Male, Male Female Female, Female Male Female, or Female Male Male. This letter shows that this over-representation is likely due to a statistical artifact,  arising from parent preferences for mixed-sex children. As noticed in \cite{angrist1998children} and supported by the data in \cite{wang2025sex}, parents are more likely to have a third child if their first two children are of the same sex. 
We show mathematically and statistically that mixed-sex preferences lead to the over-representation of patterns like Male Male Male and Female Female Female. In fact, the patterns seen in the Nurses Health Study are perfectly consistent with sex at birth being a random coin toss.

\section{Introduction} 

\cite{wang2025sex} consider the statistics of children's sex at birth in families with 2 or more children. Conditioning on families with more than 2 children, they recognize that the observed clustering of all same-sex siblings diverges from what one would expect if sex followed a binomial distribution. However, it is well known that families whose first 2 children have the same sex are more likely to have a third child (\cite{angrist1998children}). In this letter, we demonstrate using mathematical statistical arguments that in the presence of parents' mixed-sex preferences, if one conditions on families having at least 3 children, and even if a child's sex at birth is always a random coin toss, the \emph{conditional} probability of the first two or three children having the same sex is greater than could be expected under a binomial distribution. As such, the findings from \cite{wang2025sex} do not contradict the possibility that children's sex is completely random, but rather demonstrate selection bias due to conditioning on post-baseline information (number of children in a family).

\section{Notation and assumptions; families with two same-sex first children are more likely to have a third child}\label{Section:assumptions}

First, consider only the first 3 children in a family.  Write $p_D$ for the probability that a family of 2 with mixed-sex children will have a third child, and $p_S$ for the probability that a family of 2 with same-sex children will have a third child. 
Write MMM for a family having at least 3 children, with the first 3 of them male, FMM for a family having at least 3 children, with the first of them female and the next 2 of them male, etc. Write $N$ for a family's number of children. 

The hypothesis presumably falsified in \cite{wang2025sex} is the following:

\begin{assumption}\label{cointoss} (random coin toss).
Biological sex is a coin toss, and each subsequent child has the same probability of being male, $p_M$.
\end{assumption}

Write $p_F=1-p_M$ for the probability of female sex. 
\cite{wang2025sex} excluded families with twins, triplets, etc. Similarly, this letter simplifies the derivations and assumes that all births are singletons. (In practice, excluding families with first-born twins or triplets can be seen as selection on baseline information and so would not bias the analysis; excluding families with subsequent twins or triplets could bias the analysis. We provide extended results in the Appendix that do not suffer from such bias.) 

The following has long been known to be true (\cite{angrist1998children}), and also holds in the dataset presented in \cite{wang2025sex}:
\begin{assumption}\label{bigger} (parents' mixed-sex preferences). $p_S>p_D$.
\end{assumption}
From the dataset, $\hat{p}_S=0.426$ and $\hat{p}_D=0.354$, where hats represent estimates of quantities.

The next two assumptions are not required for the main results in this letter, but they simplify the inflation factor (deviation from a binomial distribution) due to conditioning on families with 3 or more children compared to a binomial in Corollary~\ref{cormain}.

\begin{assumption}\label{secondchild} The sex of the first child does not predict whether a family has a second child.
\end{assumption}

\begin{assumption}\label{thirdchild} Only same-sex of the first two children increases the probability of a family having a third child and the increase is the same regardless of whether the first two are both male or both female.
\end{assumption}

\section{Theory}\label{theory}

\subsection{Primary analysis conditioning on having at least 3 children}\label{primary}

First, consider the sexes of the first 3 children only, within families with at least 3 children.

\begin{theorem}\label{main}
Under only random coin toss Assumption~\ref{cointoss},
\begin{eqnarray*}
P\left(\text{MMM or FFF}\mid N\geq 3\right)
&=&p_M^3 \frac{P\left(N\geq 2\mid \text{M}\right)}{P\left(N\geq 2\right)}\frac{P\left(N\geq 3\mid N\geq 2, \text{MM}\right)}{P\left(N\geq 3\mid N\geq 2\right)}+\\
&&+
p_F^3 \frac{P\left(N\geq 2\mid \text{F}\right)}{P\left(N\geq 2\right)}\frac{P\left(N\geq 3\mid N\geq 2, \text{FF}\right)}{P\left(N\geq 3\mid N\geq 2\right)}.
\end{eqnarray*}
The first male-correction factor simplifies as
\begin{equation*}
 \frac{P\left(N\geq 2\mid \text{M}\right)}{P\left(N\geq 2\right)}=\frac{P\left(\text{M}\mid N\geq 2\right)}{p_M}
 \end{equation*}
 and the first female-correction factor simplifies as
\begin{equation*}
\frac{P\left(N\geq 2\mid \text{F}\right)}{P\left(N\geq 2\right)}=\frac{P\left(\text{F}\mid N\geq 2\right)}{p_F}.     \end{equation*}
\end{theorem}
Proofs of all theoretical results are in the Appendix. Notice that Theorem~\ref{main} leads to two correction factors for each of the terms for male and female births. The first correction factors are for how the probability of a second child depends on the sex of the first child. The second correction factors are for how the probability of a third child depends on the sex of the first two children.

Theorem~\ref{main} implies that if biological sex is a coin toss, the probability that the first 3 children have the same sex in a family with 3 or more does not come from a binomial distribution with probability $p_M^3+p_F^3$. The deviation arises from the conditioning on having 3 or more children.  

Under the additional assumptions, Corollary~\ref{cormain} below simplifies the expression of the deviation from a binomial distribution, which depends in particular on $p_S/p_D$: how much the probability of having a third child is increased when the first two children have the same sex.
\begin{cor}\label{cormain}
Under random coin toss Assumption~\ref{cointoss} and Assumptions~\ref{secondchild} and \ref{thirdchild},
\begin{equation*}
P\left(\text{MMM or FFF}\mid N\geq 3\right)=\left(p_M^3+p_F^3\right)\frac{1}{2 p_Fp_M p_D/p_S +p_F^2+p_M^2},    
\end{equation*}
with the inflation factor
\begin{equation*}
\frac{1}{2 p_Fp_M p_D/p_S +p_F^2+p_M^2}>1    
\end{equation*}
if additionally parents' mixed-sex preferences Assumption~\ref{bigger} holds.
\end{cor}

\subsection{Conditioning on having exactly 3 children}
One analysis in \cite{wang2025sex} conditioned on having a family with exactly three children ($N=3$) and investigated the proportion of FFF or MMM. Given Theorem~\ref{main},  under only random coin toss Assumption~\ref{cointoss},
\begin{eqnarray*}
\lefteqn{P\left(\text{MMM}\mid N=3\right)}\\
&=&\frac{P\left(\text{MMM and } N=3\right)}{P\left(N= 3\right)}\\
&=&\frac{P\left(N=3\mid \text{MMM and } N\geq 3\right)}{P\left(N=3\mid N\geq 3\right)}\frac{P\left(\text{MMM and } N\geq 3\right)}{P\left(N\geq 3\right)}\\
&=&p_M^3 \frac{P\left(N\geq 2\mid M\right)}{P\left(N\geq 2\right)}\frac{P\left(N\geq 3\mid N\geq 2, MM\right)}{P\left(N\geq 3\mid N\geq 2\right)}\frac{P\left(N=3\mid N\geq 3, MMM\right)}{P\left(N=3\mid N\geq 3\right)},
\end{eqnarray*}
where the last line uses Theorem~\ref{main}.
Thus, conditioning on $N=3$ adds factors
\begin{equation*}\frac{P\left(N=3\mid N\geq 3, \text{MMM}\right)}{P\left(N=3\mid N\geq 3\right)}
\;\;\;\;\text{  and  }\;\;\;\;\frac{P\left(N=3\mid N\geq 3, \text{FFF}\right)}{P\left(N=3\mid N\geq 3\right)}
\end{equation*}
to the terms in Theorem~\ref{main}.  Both of these factors are $<1$ if families where the first 3  children have the same sex are more likely to have a fourth child. Thus, conditioning on having exactly 3 children can be expected to \emph{attenuate} the positive bias caused by the first two correction factors.

\subsection{Sensitivity analysis exclusing last child's sex}

In their ``most conservative'' analysis, \cite{wang2025sex} excluded the last birth in the family. We now consider only the sex of the first 2 children in families with at least 3 children. Let MM? and FF? represent the sexes of the first two children with the sex of the third child ignored.

\begin{theorem}\label{secondary}
Under Assumptions~\ref{cointoss}, \ref{secondchild}, and~\ref{thirdchild},
\begin{equation*}
P\left(\text{MM? or FF?}\mid N\geq 3\right)
=\frac{1}{2 p_Fp_M p_D/p_S +p_F^2+p_M^2}\bigl(p_M^2+p_F^2\bigr).
\end{equation*}
Under Assumption~\ref{bigger}, the inflation factor
\begin{equation*}
\frac{1}{2 p_Fp_M p_D/p_S +p_F^2+p_M^2}>1.
\end{equation*}
\end{theorem}
The inflation factor in Theorem~\ref{secondary} is the same as in Theorem~\ref{main}. This was to be expected, since if we only consider the first 3 children, under random coin toss Assumption~\ref{cointoss}, the sex of the third child indeed follows a binomial distribution. Because $\bigl(p_M^2+p_F^2\bigr)>\bigl(p_M^3+p_F^3\bigr)$, the bias is \emph{larger} on the probability scale when excluding the third child. That explains why \cite{wang2025sex} find a larger deviation from the binomial distribution in their sensitivity analysis.\\

\section{Empirical results}\label{results}
We use Theorem~\ref{main} to show that the data from \cite{wang2025sex} are actually consistent with random coin toss Assumption~\ref{cointoss}. 
Focusing on the primary analysis from \cite{wang2025sex}, we use Theorem~\ref{main} instead of Corollary~\ref{cormain}, because Theorem~\ref{main} does not rely on Assumptions~\ref{secondchild} and~\ref{thirdchild}. Using the data from \cite{wang2025sex}, we estimated the two correction factors that are about decision making around having children in families: the first correction factors for males and for females are 0.522/0.516=1.0117 and 0.478/0.484=0.9876, respectively, and the second correction factors for males and for females are 0.428/0.390=1.0989 and 0.423/0.390=1.0856, respectively. Thus, in these data most of the selection bias is due to the second correction factors -- primarily the sex of the first two children affecting the probability of having a third child.

We also used the data to estimate $\hat{p}_M=0.5164$ and $\hat{p}_F=0.4836$, using only 2nd and subsequent births  since not all first births are reported.

Plugging these estimated correction factors and sex probabilities into Theorem~\ref{main}, if biological sex is like a coin toss (Assumption~\ref{cointoss}), 
\begin{eqnarray*}
\hat{P}\left(\text{MMM or FFF}\mid N\geq 3\right)
&=&\hat{p}_M^3 \frac{\hat{P}\left(\text{M}\mid N\geq 2\right)}{\hat{p}_M}\frac{\hat{P}\left(N\geq 3\mid N\geq 2, \text{MM}\right)}{\hat{P}\left(N\geq 3\mid N\geq 2\right)}+\\
&&+
\hat{p}_F^3 \frac{\hat{P}\left(\text{F}\mid N\geq 2\right)}{\hat{p}_F}\frac{\hat{P}\left(N\geq 3\mid N\geq 2, \text{FF}\right)}{\hat{P}\left(N\geq 3\mid N\geq 2\right)}\\
&=&0.5164^3\cdot 1.0117 \cdot 1.0989+\\
&&+0.4836^3\cdot 0.9876 \cdot 1.0856\\
&=&0.2743.
\end{eqnarray*}
Comparing this prediction to $\hat{p}_M^3+\hat{p}_F^3=0.2508$ under a binomial distribution, the estimated inflation factor equals 1.094: a 9.4\% increase.

In comparison, in the data reported in \cite{wang2025sex}, the proportion of MMM or FFF among families with at least 3 children was 0.2751. The p-value for the Chi-square test for whether the underlying proportion of MMM or FFF in the data is equal to 0.2743 equals p=0.786; there is hardly any deviation compared to what is expected based on Theorem~\ref{main}. 

Details of two additional hypothesis tests that consider the exclusion of families with twins, triplets, etcetera, are in the Appendix. The p-value of the additional hypothesis tests equal p=0.817, p=0.347, and p=0.641, respectively.


\section{Simulation study results}

We use simulated data in order to illustrate what we may expect from data arising under Assumptions~\ref{cointoss}, \ref{bigger}, \ref{secondchild}, and \ref{thirdchild}. In a population of families with at least 2 children, we generated random sex at each birth, with probability $p_M=0.5164$ (in the data, $\hat{p}_M=0.5164$, see Section~\ref{results}), and probability of having a 3rd child given that the first two have different sexes $p_D=0.354$ (in the data, $\hat{p}_D=0.354$). We generated 1,000 datasets of sample size 58,007, for each of 100 values of $p_S/p_D$  between 1 (no sex preference) and 1.5 (preference for mixed-sex children increases the probability of having a 3rd child by 50\%); this resulted in a total of 100,000 datasets. 

\begin{figure}\centering
\includegraphics[width=4.52in]{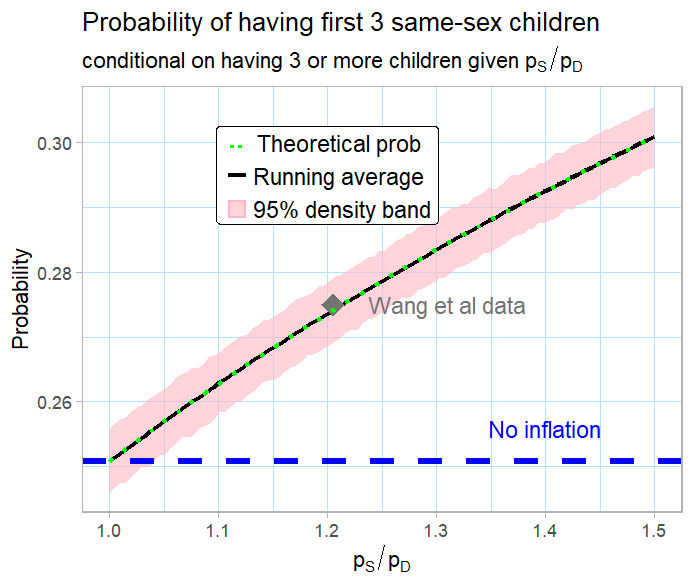}\caption{Simulation study results: loess curve and 95\% density band, plotted using $100\times1000$ datasets of sample size 58,007, over 100 values of $p_S/p_D$ between 1 and 1.5. The blue dashed line represents the probability under $p_S/p_D=1$ (no mixed-sex preference) and the grey diamond is the data-point derived from the data in \cite{wang2025sex}. The green dotted line represents the theoretical values given by Corollary~\ref{cormain}. The simulations use the data-derived estimates $\hat{p}_M, \hat{p}_F$ and $\hat{p}_D$, and vary $p_S$. $p_S/p_D$ is estimated in the data from \cite{wang2025sex} as $\hat{p}_S/\hat{p}_D=1.205$.
\label{Sim_results}}
\end{figure}
Figure~\ref{Sim_results} presents the simulation results. The graph's black line represents the  proportion of families with first three children of same sex out of the total number of families with at least three children, averaged over the datasets at each value of $p_S/p_D$. The light pink shaded region around the line represents the 95\% density band, which is the interval between the 5th and 95th percentiles of the data estimates. Thus, we would expect that 95\% of the time, data generated under our assumptions, including random coin toss Assumption~\ref{cointoss}, would have a proportion of first three children MMM or FFF out of all families with 3 or more children fall within this region.   The average simulated proportions (black line) correspond exactly with the expected probabilities (green dotted line) obtained from Corollary~\ref{cormain}. The simulations use the data-derived estimates $\hat{p}_M, \hat{p}_F$ and $\hat{p}_D$, and vary $p_S$. 
The blue dotted line represents the probability of the first 3 children having the same sex conditional on having at least 3 children in the setting where there is no mixed-sex preference ($p_S/p_D=1$). 

The grey diamond is derived directly from the data in \cite{wang2025sex}, at x-coordinate $\hat{p}_S/\hat{p}_D=1.205$ and y-coordinate $\hat{P}\left(\text{MMM or FFF}\mid N\geq 3\right)=0.2751$: the observed proportion of MMM or FFF among families with at least 3 children. Because the grey diamond falls within the 95\% density bands and in fact, almost directly on the average line, the data in \cite{wang2025sex} are perfectly consistent with random coin toss Assumption~\ref{cointoss}.

\section{Discussion}

It is well-known (\cite{angrist1998children}) that families with two same-sex children are more likely to have a third child. This letter proves how these mixed-sex preferences explain away the findings from \cite{wang2025sex}: first by mathematical derivations, then from the data provided in \cite{wang2025sex}, and then in a simulations study. We find that in fact, the data presented are entirely consistent with sex at birth being a biological coin toss.

\section*{R-code and data}
The R-code that generated the numbers in the results section is available upon request to jjlok@bu.edu. The R-code that generated the results in the simulation section is available upon request to mireille.schnitzer@umontreal.ca. The data analyzed in the results section were published in \cite{wang2025sex}.

\addcontentsline{toc}{chapter}{Bibliography}
\bibliographystyle{chicago} \bibliography{ref}

\appendix
\section{Appendix}
\label{sec:appendix}

\subsection{Proofs of theoretical results}
\noindent {\bf Proof of Theorem~\ref{main}}
\begin{eqnarray*}
\lefteqn{P\left(\text{MMM}\mid N\geq 3\right)\;\;\;\;\;\;\;\;\;\;\;\;\;\;\;\;\;\;\;\;\;\;\;\;\;\;\;\;\;\;\;\;\;\;\;\;}\\
&=&\frac{P\left(\text{MMM},N\geq 3\right)}{P\left(N\geq 3\right)}\\
&=&\frac{P\left(\text{MMM}\mid N\geq 3, \text{MM}\right)P\left(N\geq 3, \text{MM}\right)}{P\left(N\geq 3\right)}\\
&=&\frac{p_M P\left(N\geq 3\mid N\geq 2, \text{MM}\right)P\left(N\geq 2, \text{MM}\right)}{P\left(N\geq 3\right)}\\
&=&\frac{p_M P\left(N\geq 3\mid N\geq 2, \text{MM}\right)P\left(N\geq 2,\text{MM}\mid \text{M}\right)P\left(\text{M}\right)}{P\left(N\geq 3\right)}\\
&=&\frac{p_M^2 P\left(N\geq 3\mid N\geq 2, \text{MM}\right)P\left(\text{MM}\mid N\geq 2, \text{M}\right)P\left(N\geq 2\mid \text{M}\right)}{P\left(N\geq 3\right)}\\
&=&\frac{p_M^3 P\left(N\geq 3\mid N\geq 2, \text{MM}\right)P\left(N\geq 2\mid \text{M}\right)}{P\left(N\geq 3\mid N\geq 2\right)P\left(N\geq 2\right)}\\
&=&p_M^3 \frac{P\left(N\geq 3\mid N\geq 2, \text{MM}\right)}{P\left(N\geq 3\mid N\geq 2\right)}\frac{P\left(N\geq 2\mid \text{M}\right)}{P\left(N\geq 2\right)}.
\end{eqnarray*}
The third, fifth, and sixth equalities use random coin toss Assumption \ref{cointoss}.\\ $P\left(\text{FFF}\mid N\geq 3\right)$ follows similarly.

The first correction factor for families that first had two male children equals
\begin{eqnarray*}
 \frac{P\left(N\geq 2\mid \text{M}\right)}{P\left(N\geq 2\right)}
 &=& \frac{P\left(N\geq 2, \text{M}\right)}{P\left(\text{M}\right)P\left(N\geq 2\right)}\\
 &=& \frac{P\left(\text{M}\mid N\geq 2\right)P\left(N\geq 2\right)}{p_MP\left(N\geq 2\right)}\\
 &=& \frac{P\left(\text{M}\mid N\geq 2\right)}{p_M}.
\end{eqnarray*}
The first correction factor for families that first had two female children follows similarly.
$\hfill\Box$\\

\noindent {\bf Proof of Corollary~\ref{cormain}}
Under Assumption~\ref{secondchild}, the first male and female correction factors in the result of Theorem~\ref{main} are equal to 1. Under Assumptions~\ref{secondchild}, \ref{cointoss}, and~\ref{thirdchild}, the denominators of the second correction factors in the result of Theorem~\ref{main} are both equal to
\begin{eqnarray*}
P\left(N\geq 3\mid N\geq 2\right)
&=&P\left(N\geq 3\mid N\geq 2, \text{MM}\right)P\left(\text{MM}\mid N\geq 2\right)\\
&&+P\left(N\geq 3\mid N\geq 2, \text{FF}\right)P\left(FF\mid N\geq 2\right)\\
&&+P\left(N\geq 3\mid N\geq 2, \text{MF}\right)P\left(\text{MF}\mid N\geq 2\right)\\
&&+P\left(N\geq 3\mid N\geq 2, \text{FM}\right)P\left(\text{FM}\mid N\geq 2\right)\\
&=&p_SP\left(\text{MM}\mid N\geq 2\right)
    +p_SP\left(\text{FF}\mid N\geq 2\right)\\
&& +p_DP\left(\text{MF}\mid N\geq 2\right)
    +p_DP\left(\text{FM}\mid N\geq 2\right)\\
&=&p_SP\left(\text {MM or FF}\mid N\geq 2\right)+p_DP\left(\text{MF or FM}\mid N\geq 2\right)\\
&=&p_S \left(p_M^2+p_F^2\right)+2p_Dp_Mp_F.
\end{eqnarray*}
If additionally Assumption~\ref{bigger} holds ($p_D<p_S$), this denominator of the second correction term in Theorem~\ref{main} is smaller than the numerator $p_S$ (since $p_M^2+p_F^2+2p_Mp_F=1$), so that the inflation factor is $>1$.$\hfill\Box$\\

\noindent {\bf Proof of Theorem~\ref{secondary}.}
\begin{eqnarray*}
\lefteqn{P\left(\text{MM? or FF?}\mid N\geq 3\right)}\nonumber\\
&=&\frac{P\left(\text{MM? and }N\geq 3\right)+P\left(\text{FF? and }N\geq 3\right)}{P\left(N\geq 3\mid N\geq 2, D\right)P\left(D,N\geq 2\right)+P\left(N\geq 3\mid N\geq 2, S\right)P\left(S,N\geq 2\right)}.
\end{eqnarray*}
Similar to the proof of Theorem~\ref{main},
\begin{eqnarray*}
P\left(\text{MM? and }N\geq 3\right)&=&P\left(N\geq 3, \text{MM}\mid \text{MM}, N\geq 2\right)P\left(\text{MM}, N\geq 2\right)\\
&=&p_S p_M^2P\left(N\geq 2\right),
\end{eqnarray*}
with a similar expression for FF?, so that with the same reasoning as in the proof of Theorem~\ref{main},
\begin{eqnarray*}
\lefteqn{P\left(\text{MM? or FF?}\mid N\geq 3\right)}\\
&=&(p_M^2+p_F^2)\frac{p_S P\left(N\geq 2\right)}{p_DP\left(D,N\geq 2\right)+p_SP\left(S,N\geq 2\right)}\\
&=&(p_M^2+p_F^2)\frac{1}{p_D/p_SP\left(D,N\geq 2\mid N\geq 2\right)+P\left(S,N\geq 2\mid N\geq 2\right)}\\
&=&(p_M^2+p_F^2)\frac{1}{2 p_M p_F p_D/p_S+\left(p_M^2+p_F^2\right)}\\
&>&p_M^2+p_F^2,
\end{eqnarray*}
with the same inflation factor compared to assuming a binomial distribution as in Corollary~\ref{cormain}. $\hfill\Box$\\

\subsection{Two additional hypothesis tests not sensitive to excluding twins, triplets, etcetera}

Respecting the order of decision making to focus only on the sex of a child given the previous history, we tested
\begin{equation*}H_0: P\bigl(\text{same sex}\mid \text{same sex so-far and another child}\bigr)
\end{equation*}
in the data reported in \cite{wang2025sex}, starting as suggested in \cite{wang2025sex} at the third birth. Given that all of at least 2 previous children are male, combining families with 3, 4, and 5 children and conditioning on $N\geq 3, 4, 5$, respectively, the estimated probability of the next child being male is 0.515, slightly less than expected on average 0.516 (p=0.817, Chi-square test). Given that all of at least 2 previous children are female, the estimated probability of the next child being female is 0.489, slighty more than expected on average 0.484, but not significantly so (p=0.347, Chi-square test).

We also carried out a combined-sex test respecting the order of decision making, using as the null hypothesis that the overall probability of observing ``same sex'' given repeated-sex children equals $\hat{p}_F f+\hat{p}_M(1-f)=0.5014$, where $f$ is the fraction of family-instances with initial repeated female children contributing to the analysis. This test conditions the third child on the sex of the first 2 children, the fourth child on the sex of the first 3 children, and the fifth child on the sex of the first 4 children. The resulting p-value is 0.641 (Chi-square test).


\end{document}